\documentclass[aps,prd,twocolumn,showpacs]{revtex4}

\usepackage{amsmath}
\usepackage{lipsum}
\usepackage{amssymb}
\usepackage{amsthm}
\usepackage{bbold}
\usepackage{dcolumn}
\usepackage{epsfig}
\usepackage{graphics}
\usepackage{graphicx}
\usepackage{color}
\usepackage{xspace}
\usepackage{cancel}
\usepackage{IEEEtrantools}
\usepackage[utf8]{inputenc}

\usepackage{hyperref}
\definecolor{darkred}{rgb}{0.5,0,0}
\definecolor{darkgreen}{rgb}{0,0.5,0}
\definecolor{darkblue}{rgb}{0,0,0.5}
\hypersetup{ colorlinks,
linkcolor=darkblue,
filecolor=darkgreen,
urlcolor=darkred,
citecolor=darkblue }

\newcommand{\MeV}{\text{MeV}} 
\newcommand{\GeV}{\text{GeV}} 

\newcommand{\eqn}[1]{Eq.~(\ref{#1})}

\newcommand{\tab}[1]{Table~\ref{#1}}
\newcommand{\sect}[1]{Section~\ref{#1}}

\newcommand{\df}[1]{\hspace{-0.5em}\ensuremath{\frac{\mathrm{d}^{4}#1}{(2\pi)^{4}}}\,}

\begin{document}

\title{Heavy-light mesons spectra in a contact interaction }

\author{Marco A. Bedolla$^{1,2}$, E. Santopinto$^{2}$, and L. X. Guti\'{e}rrez-Guerrero$^{3}$}
\affiliation{
$^1$ Instituto de F\'{i}sica y Matem\'aticas, Universidad Michoacana de San
Nicol\'as Hidalgo, Edificio C-3, Ciudad Universitaria,
Morelia, Michoac\'an 58040, M\'exico.}
\affiliation{
$^2$ Istituto Nazionale di Fisica Nucleare (INFN), Sezione di Genova, via Dodecaneso 33, 16146 Genova, Italia.}
\affiliation{
$^3$ CONACyT-Mesoamerican Centre for Theoretical Physics,
Universidad Aut\'onoma de Chiapas, Carretera Zapata Km. 4,
Real del Bosque (Ter\'an), Tuxtla Guti\'errez 29040, Chiapas, M\'exico.}

\date{\today}

\begin{abstract}

  We present the spectrum and decay constants of heavy-light mesons in four different channels: pseudo-scalar, vector,
  scalar and axial vector.  We extend the framework for our previous analysis in a unified
  symmetry-preserving Schwinger-Dyson equations (SDE) treatment of a vector$\times$vector
  contact interaction. Despite the simplicity of our model, the results found for the meson masses are in good agreement experimental data and earlier model calculations based upon Schwinger-Dyson and Bethe-Salpeter equations (BSEs) involving  sophisticated interaction kernels.  
\end{abstract}

\keywords{Heavy quarkonium, mass spectrum, radial excitations,
  Bethe-Salpeter equation, confinement, dynamical chiral symmetry breaking, Schwinger-Dyson
  equations, contact interaction}

\maketitle

\date{\today}

\section{\label{sec:intro}Introduction}

Comprehensive studies on heavy-light mesons are an important piece to completely understand QCD. Due to the presence of different quark masses inside these states, they might depict a bridge between the regime of chiral dynamics and heavy quarks symmetries~\cite{Hilger:2008jg,Hilger:2011cq}. However, due to their complexity, an extensive study of these systems has been neglected by most of the works in theoretical hadron physics with the Schwinger-Dyson and Bethe-Salpeter Equations (SDBSE) approach~\cite{Jain:1993qh,Maris:1997tm,Maris:2000sk,Burden:2002ps,Maris:2005tt,Maris:2006ea,Bhagwat:2007rj,Eichmann:2008ef,Blank:2010sn,Mader:2011zf,Popovici:2014pha,Sanchis-Alepuz:2014sca,Eichmann:2015cra}.

We use a contact interaction (CI) model that appeared as an alternative to conduct exploratory studies on QCD  within the SDBSE approach~\cite{GutierrezGuerrero:2010md,Roberts:2010rn,Roberts:2011cf,Roberts:2011wy,Chen:2012qr}. In this model, quarks interact not via mass-less vector-boson exchanges, but instead through a symmetry preserving vector-vector contact interaction. This interaction is embedded within the SDBSE approach in the rainbow-ladder approximation, implement confinement through a proper time regularization scheme~\cite{Ebert:1996vx}. 

This interaction provides a good description of light ground and excited states meson and baryon spectra~\cite{GutierrezGuerrero:2010md,Roberts:2010rn,Roberts:2011cf,Roberts:2011wy,Chen:2012qr}, and heavy quarkonia~\cite{Bedolla:2015mpa,Raya:2017ggu}. The results derived from the CI model are quantitatively comparable to those obtained using sophisticated QCD model interactions,~\cite{Maris:2006ea,Cloet:2007pi,Eichmann:2008ae,Bashir:2012fs}.

In this work, we present one the first steps towards the study of heavy-light systems by the contact interaction model. Our results are a direct application of this model in the heavy sector~\cite{Bedolla:2015mpa,Bedolla:2016yxq,Raya:2017ggu}. We calculate the heavy-light mesons spectra and compare them experimental data and other covariant model setups~\cite{Nguyen:2010yh,Fischer:2014cfa,Gomez-Rocha:2016cji,Hilger:2017jti}.

This paper is organized as follows: in~\sect{sec:ci-mod} we give the minimum details necessary to
the SDBSE approach to study mesons, we describe briefly the contact interaction in the rainbow-ladder approximation. Furthermore, we present an unified CI model framework to study light, heavy and heavy-light mesons.  
In~\sect{sec:massspectrum}, we depict our results for the mass spectrum of heavy-light mesons that can be calculated with our scheme. Finally, in \sect{sec:conclusions}, we state our conclusions.

\section{\label{sec:ci-mod} DSBSE Approach in a Contact interaction model}

Since this work is a direct application of the unified CI model presented in Ref.~\cite{Raya:2017ggu}, we only sketch the basic formulae. The complete description of this model is found in Refs.~\cite{Chen:2012qr,Bedolla:2015mpa,Raya:2017ggu}. 

\subsection{Contact Interaction Model}

The DSBSE approach solves the bound-state problem in terms of their building blocks (quarks) and their interactions with gluons. In order to solve the meson bound state equation, we need to know the quark propagator, the gluon propagator and the quark-gluon interaction. In the contact interaction model, we assume that the quark-gluon interaction is led by symmetry-preserving vector$\times$vector contact interaction; here, quarks are attached through the interaction defined as 
\begin{eqnarray}
\label{eqn:contact_interaction}
g^{2}D_{\mu \nu}(k)&=&\frac{4\pi\alpha_{\mathrm{IR}}}{m_g^2}\delta_{\mu \nu} \equiv
\frac{1}{m_{G}^{2}}\delta_{\mu\nu}, \\
\label{eqn:quark_gluon_vertex_rl}
\Gamma^{a}_{\mu}(p,q)&=&\frac{\lambda^{a}}{2}\gamma_{\mu},
\end{eqnarray}
\noindent where $m_g=800\,\MeV$ is a gluon mass scale generated dynamically in QCD~\cite{Boucaud:2011ug}, and $\alpha_{\mathrm{IR}}$  is the CI model parameter, which can be interpreted as the interaction strength in the infrared~\cite{Binosi:2016nme,Deur:2016tte}.

With this interaction, we obtain a constant mass function. Because the integrals we need to solve are divergent, we adopt the proper time regularization scheme~\cite{Ebert:1996vx}, and we introduce the parameters $\tau_{\mathrm{IR}}$ and $\tau_{\mathrm{UV}}$ as infrared and ultraviolet
regulators, respectively. A nonzero value for  $\tau_{\mathrm{IR}}\equiv 1/\Lambda_{\mathrm{IR}}$ implements 
confinement~\cite{Roberts:2007ji}. Since the CI is nonrenormalizable theory, 
$\tau_{\mathrm{UV}}\equiv 1/\Lambda_{\mathrm{UV}}$ becomes part of the model and therefore sets the scale for
all dimensional quantities.

The next step in our setup is to solve the Bethe-Salpeter Equation (BSE) to obtain. The solution of the BSE, the Bethe-Salpeter Amplitudes (BSA), is combined with the quark propagator in the interaction kernel. Namely, the homogeneous BSE in an explicit $J^{PC}$ channel is~\cite{Gross:1993zj,Salpeter:1951sz,GellMann:1951rw},
\begin{equation}
\label{eqn:bse}
\left[\Gamma_{H}(p;P)\right]_{tu}=
\int\df{q}K_{tu;rs}(p,q;P)\chi(q;P)_{sr},
\end{equation}
\noindent where $\chi(q;P)=S_{f}(q_{+})\Gamma_{H}(q;P)S_{g}(q_{-})$ is the Bethe-Salpeter wave-function; $q_{+}=q+\eta P$,
$q_{-}=q-(1-\eta)P$; $\eta \in [0,1]$ is a momentum-sharing parameter, $p$ ($P$) is the relative
(total) momentum of the quark-antiquark system; $S_{f}$ is the $f$-flavor dressed-quark propagator; 
$\Gamma_{H}(p;P)$ is the meson Bethe-Salpeter amplitude (BSA), where $H$ specifies the quantum
numbers and flavor content of the meson; $r,s,t$, and $u$ represent color, Dirac and flavor indices; and $K(p,q;P)$ is the quark-antiquark scattering kernel.

\subsection{\label{sec:unifiedCImodel} CI model running coupling}

\begin{table}[h]
\begin{center}
\begin{tabular}{lllll}
\hline \hline
 quark & $\hat{\alpha}_{\mathrm {IR}}\;[\GeV^{-2}]$ & $\Lambda_{\mathrm {UV}}\;[\GeV] $ & $\alpha$ & Ratio \\
\hline
$u,d,s$ & 4.565 & 0.905 & 3.739 & 1 \\
$c$     & 0.228 & 2.400 & 1.547 & 0.414 \\
$b$     & 0.035 & 6.400 & 1.496 & 0.400 \\
\hline \hline
\end{tabular}
\caption{\label{tab:rc} Dimensionless coupling constant $\alpha=\hat{\alpha}_{\mathrm IR}\Lambda_{\mathrm UV}^{2}$,
  where $\hat{\alpha}_{\mathrm {IR}}=\alpha_{\mathrm {IR}}/m_{g}^{2}$, for the contact interaction, extracted from a
  best-fit to data, as explained in Ref.~\cite{Raya:2017ggu}. Fixed parameters are $m_{g}=0.8\,\GeV$ and
  $\Lambda_{\mathrm {IR}}=0.24\,\GeV$.}
\end{center}
\end{table}
%
%
In~\tab{tab:rc} we present the set of parameters used in previous light-mesons and heavy-quarkonia studies~\cite{Raya:2017ggu}. When studying the heavy sector, a change in the model parameters has to be done: an increase in the ultraviolet regulator, and a reduction in the coupling strength~\cite{Bedolla:2015mpa}. With these parameters, we defined a dimensionless coupling $\alpha$ guided by~\cite{Farias:2005cr,Farias:2006cs}
\begin{equation}
\alpha=\frac{\alpha_{\mathrm{IR}}}{m_{g}^{2}}\Lambda_{\mathrm{UV}}^2.\label{eqn:dimensionless_alpha}
\end{equation}
\noindent The drop in $\alpha$, related to its value in the light-quarks sector, can be
read off from the last column of \tab{tab:rc}. Indeed, $\alpha$ is reduced by a factor of
$2.1-2.3$ on going from the light to the heavy sector, instead of the apparent large factors
listed in the second column of~\tab{tab:rc}.

Moreover, as a reminiscent of the running coupling QCD with the momentum scale at which it is measured, an inverse logarithmic curve can fit reasonably well the functional dependence of $\alpha(\Lambda_{\mathrm {UV}})$. The fit reads
\begin{equation}
\label{eqn:logaritmicfit} \alpha(\Lambda_{\mathrm{UV}})=a\ln^{-1}\left(\Lambda_{\mathrm {UV}}/\Lambda_0\right) \,,
\end{equation}
where $a=0.923$ and $\Lambda_0=0.357$ ~\cite{Raya:2017ggu}. With this fit, we can recover the value of the strength coupling $\alpha$ once given a value of $\Lambda_{\mathrm {UV}}$.

\section{\label{sec:massspectrum} Heavy-light Mesons}

In order to calculate mass spectra and decay constants, we follow the expressions found in Ref.~\cite{Bedolla:2015mpa}. In our approach, to solve these unequal mass systems, we find $\Lambda_{\mathrm{UV}}$ and $\alpha_{\mathrm{IR}}$ through Eqs.~\ref{eqn:dimensionless_alpha}-\ref{eqn:logaritmicfit} in the Bethe-Salpeter equation fitted to the pseudoscalar meson experimental mass value. Later, with these parameters, we calculate the other meson masses, the pseudoscalar and vector decay constants. We use the quarks dressed masses calculated in our previous works listed in \tab{tab:DressedMasses}, and we consider those mesons with quark $s$ with the same parameters as $u$ and $d$ to be consistent with \tab{tab:rc}. We  expect the $\Lambda_{\mathrm{UV}}$ parameter within the domain $(0.905, 2.4)$ for charmed mesons, $(0.905, 6.4)$ for $B$ and $B_s$ mesons, and $(2.4, 6.4)$ for $B_c$ mesons.
\begin{table}[h]
\begin{center}
\begin{tabular}{lllll}
\hline \hline
 quark & $u, d$ & $s $ & $c$ & $b$ \\
\hline
Mass & 367 & 533 & 1482 & 4710 \\
\hline \hline
\end{tabular}
\caption{\label{tab:DressedMasses} Dressed quark masses (in Mev) generated dynamically for the CI model defined in \eqn{eqn:contact_interaction}  using the parameters enlisted in \tab{tab:rc}.}
\end{center}
\end{table}
%
 Additionally, we multiply the coupling of scalar and axial-vector charmed mesons by two distinct spin-orbit(SO) parameters,
\begin{equation}
g_{\mathrm {SO}}^{0^{+}}=0.32,\quad g_{\mathrm {SO}}^{1^{+}}=0.25\,. \label{eqn:twoSOs}
\end{equation}
With these values, it is possible to match the mass splitting $m_{a_1}-m_{\rho}=0.45$GeV and $m_{\sigma}-\rho=0.29$GeV, the same obtained by refined Bethe-Salpeter kernels, as described in Ref.~\cite{Lu:2017cln}. Morever, for the calculation of bottom mesons we divide the SO parameters by an additional factor of 3~\cite{Bedolla:2015mpa}.

We present tables comparing different values computed with the CI model, other SDBSE models results and experimental data when they are available.
\subsection{\label{sec:D-Mesons} $D$ and $D_s$ Mesons}
\begin{table}[h]
\begin{center}
\begin{tabular}{lcccc}
\hline \hline
    & \multicolumn{4}{c}{masses and decay constants [MeV]}  \\
    & \multicolumn{2}{c}{$(m,f)$} & \multicolumn{2}{c}{$m$ } \\
\hline
&$ D$ & $D^*$ & $D_{0}$ & $D_{1}$
\\
\hline
Experiment~\cite{Olive:2016xmw}$^*$ & (1864, 149) & (2010, 196) & 2318 & 2420  \\
CI-model & (1864, 267) & (2068, 162) & 2300 & 2386 \\
CI-subtr~\cite{Serna:2017nlr} & (1869, 146) & (2011, 169) & $\cdots$ & $\cdots$ \\
NST1~\cite{Nguyen:2010yh} & (1850, 108) & (2040, 113)  & $\cdots$ & $\cdots$ \\
NST2~\cite{Nguyen:2010yh} & (1880, 183) & $\cdots$ & $\cdots$ & $\cdots$ \\
REBM~\cite{Rojas:2014aka} & (2115, 144) & $\cdots$ & $\cdots$ & $\cdots$ \\
HGKL1~\cite{Hilger:2017jti} & (1868, 228) & $\cdots$  & $\cdots$ & $\cdots$ \\
HGKL2~\cite{Hilger:2017jti} & (1869, 678) & $\cdots$ &$\cdots$ & $\cdots$ \\
\hline \hline
   & \multicolumn{3}{c}{amplitudes}  \\
\hline\hline
$E_H$ & 4.000 & 0.944 & 0.313 & 0.159 \\
$F_H$ & 0.191 & $\cdots$  & $\cdots$ & $\cdots$\\
\hline\hline
\end{tabular}
\caption{\label{tab:mcn_all_opt} Mass spectrum of ground-state $D$ mesons.
  Our results were obtained with the best-fit parameter set: 
  $\alpha_{\mathrm{IR}}= 0.93\pi/4.528$ and $\Lambda_{\text{UV}}= 1.532\,\GeV$.
   $^*$The decay constants are from a Lattice calculation \cite{Becirevic:2012ti}.}
\end{center}
\end{table}
In order to calculate the $D$ and $D_s$ mesons spectra, we found $\hat\alpha_{\mathrm{IR}}= 0.645$ and $\Lambda_{\text{UV}}= 1.532$, which are between the light and charm region according to the second and third column of \tab{tab:rc}. 
Our results are enlisted in \tab{tab:mcn_all_opt}. Here, we compare our spectrum of $D$-mesons with experimental data and other models predictions. These results are in excellent agreement with other data. However, decay constants predictions with different SDBSE methods do not agree with experimental values. Moreover, its value in different studies with an slightly different models, present results completely different, see Refs.~\cite{Nguyen:2010yh,Hilger:2017jti}. This inconvenient is mainly a result of using the rainbow-ladder approximation.

%
\begin{table}[h]
\begin{center}
\begin{tabular}{lcccc}
\hline \hline
    & \multicolumn{4}{c}{masses and decay constants [MeV]}  \\
       & \multicolumn{2}{c}{$(m,f)$} & \multicolumn{2}{c}{$m$ } \\
\hline
& $D_s $ & $D^*_s $ & $D_{s1} $ & $D_{s1} $
\\
\hline
Experiment~\cite{Olive:2016xmw}$^*$ & (1968, 176) & (2112, 227) & 2317 & 2459  \\
CI-model & (1949, 228) & (2157, 164) & 2408 & 2487 \\
CI-subtr~\cite{Serna:2017nlr} & (1977, 169)	 & (2098, 195) & $\cdots$ & $\cdots$ \\
NST1~\cite{Nguyen:2010yh} & (1970, 139) & (2170, 180)  & $\cdots$ & $\cdots$ \\
NST2~\cite{Nguyen:2010yh} & (1900, 194) & $\cdots$ & $\cdots$ & $\cdots$ \\
REBM~\cite{Rojas:2014aka} & (2130, 176) & $\cdots$ & $\cdots$ & $\cdots$ \\
HGKL1~\cite{Hilger:2017jti} & (1872, 190) & (2175, 125) & 2265 & 2354 \\
HGKL2~\cite{Hilger:2017jti} & (1802, 208) & (2011, $\cdots$) & 2211 & $\cdots$ \\
\hline \hline
   & \multicolumn{3}{c}{amplitudes}  \\
\hline\hline
$E_H$ & 3.477 & 1.000 & 0.302 & 0.155 \\
$F_H$ & 0.163 & $\cdots$  & $\cdots$ & $\cdots$\\
\hline\hline
\end{tabular}
\caption{\label{tab:mcs_all_opt} Mass spectrum of ground-state $D_s$  mesons.
  Our results were obtained with the best-fit parameter set:
  $\alpha_{\mathrm{IR}}= 0.93\pi/4.528$ and $\Lambda_{\text{UV}}= 1.532\,\GeV$.
   $^*$The decay constants are from a Lattice calculation~\cite{Becirevic:2012ti}.}
\end{center}
\end{table}

In \tab{tab:mcs_all_opt} we compare the spectrum of $D_s$ mesons. Again, our predictions are in excellent agreement with experimental values and other SDBSE calculations. A reduction in the strange dressed mass improve slightly the predictions~\cite{Serna:2017nlr}, though our intention is to maintain coherence with previous CI model studies~\cite{Chen:2012qr}.

\subsection{\label{sec:Bottom-Mesons} $B$ and $B_s$ mesons}

%
\begin{table}[h]
\begin{center}
\begin{tabular}{lcccc}
\hline \hline
    & \multicolumn{4}{c}{masses and decay constants [GeV]}  \\
    & \multicolumn{2}{c}{$(m,f)$} & \multicolumn{2}{c}{$m$ } \\
\hline
& $B$ & $B^*$ & $B_{0}$ & $B_{1}$
\\
\hline
Experiment~\cite{Olive:2016xmw}$^*$ & (5279, 131) & (5325, 123) & $\cdots$ & 5725  \\
CI-model & (5279, $\cdots$) & (5325, 145)  & 5610 & 5661 \\
NST1~\cite{Nguyen:2010yh} & (5270, 74) & (5150, 187)  & $\cdots$ & $\cdots$ \\
NST2~\cite{Nguyen:2010yh} & (5150, 187) & $\cdots$ & $\cdots$ & $\cdots$ \\
\hline \hline
   & \multicolumn{3}{c}{amplitudes}  \\
\hline\hline
$E_H$ & 0.773 & 0.822 & 0.076 & 0.041 \\
$F_H$ & 0.004 & $\cdots$  &$\cdots$ & $\cdots$\\
\hline\hline
\end{tabular}
\caption{\label{tab:mbn_all_opt} Mass spectrum of ground-state of $B$  mesons.
  Our results were obtained with the best-fit parameter set:
  $\alpha_{\mathrm{IR}}= 0.93\pi/11.969$ and $\Lambda_{\text{UV}}= 2.223\,\GeV$.
  $^*$The decay constants are from a recent Lattice calculation \cite{Colquhoun:2015oha}.}
\end{center}
\end{table}
In order to calculate the $B$ and $B_s$ mesons spectra, we found $\hat\alpha_{IR}= 0.244$ and $\Lambda_{\text{UV}}= 2.223$, which are between the light and bottom region according to the second and third column of \tab{tab:rc}.
\tab{tab:mbn_all_opt} provides the CI model predicted $B$ mesons mass spectrum. We notice that the mass spectrum agrees perfectly with experimental data and other model results. Additionally, thanks to the model simplicity, we present a value for the $B_0$ mass, where other models still struggle to calculate that value, and neither there are experimental data. Because the mass difference between the experimental and our CI-model value for $B_{1}$ is 64 MeV, we expect that our $B_{0}$ mass value to be around 60 MeV below a future experimental value.

Despite our good predictions for the mass spectrum, we could not calculate any pseudoscalar decay constant. We recycled the expressions found in Ref.~{Bedolla:2015mpa} and we obtain imaginary numbers, however, we could calculate the vector decay constants. We hope that this picture will be improved in future works beyond rainbow-ladder truncation.

%
\begin{table}[h]
\begin{center}
\begin{tabular}{lcccc}
\hline \hline
    & \multicolumn{4}{c}{masses and decay constants [MeV]}  \\
       & \multicolumn{2}{c}{$(m,f)$} & \multicolumn{2}{c}{$m$ } \\
\hline
& $B_s$ & $B^*_s$ & $B_{s0}$ & $B_{s1}$
\\
\hline
Experiment~\cite{Olive:2016xmw}$^*$ & (5366, 158) & (5415, 150) & $\cdots$ & 5828  \\
CI-model & (5364, $\cdots$) & (5413, 148) & 5701 & 5747 \\
NST1~\cite{Nguyen:2010yh} & (5380, 101) & (5420, 113)  & $\cdots$ & $\cdots$ \\
NST2~\cite{Nguyen:2010yh} & (4750, 115) & $\cdots$ & $\cdots$ & $\cdots$ \\
\hline \hline
   & \multicolumn{3}{c}{amplitudes}  \\
\hline\hline
$E_H$ & 0.974 & 0.855 & 0.073 & 0.039 \\
$F_H$ & 0.005 & $\cdots$  & $\cdots$ & $\cdots$\\
\hline\hline
\end{tabular}
\caption{\label{tab:mbs_all_opt} Mass spectrum of ground-state $B_s$  mesons.
  Our results were obtained with the best-fit parameter set: 
  $\alpha_{\mathrm{IR}}= 0.93\pi/11.969$ and $\Lambda_{\text{UV}}= 2.223\,\GeV$. $^*$The decay constants are from a recent Lattice calculation \cite{Colquhoun:2015oha}.}
\end{center}
\end{table}
%
\tab{tab:mbs_all_opt} displays the CI model predictions for the $B_s$ mesons mass spectrum. One more time, we realize that the mass spectrum agrees perfectly with experimental data and other models results when they are available. Because the mass difference between the experimental and our CI-model value for $B_{s1}$ is 78 MeV, we expect that our $B_{s0}$ mass value to be around 75 MeV below a future experimental value.

\subsection{\label{sec:Bc-Mesons} $B_c$ mesons}
%
\begin{table}[h]
\begin{center}
\begin{tabular}{lcccc}
\hline \hline
    & \multicolumn{4}{c}{masses and decay constants [MeV]}  \\
       & \multicolumn{2}{c}{$(m,f)$} & \multicolumn{2}{c}{$m$ } \\
\hline
& $B_c$ & $B^*_c$ & $B_{c0}$ & $B_{c1}$
\\
\hline
Experiment~\cite{Olive:2016xmw}$^*$ & (6275, 302) & ($\cdots$, 298)& $\cdots$ & $\cdots$  \\
CI-model & (6275, $\cdots$) & (6308, 203) & 6490 & 6518 \\
NST1~\cite{Nguyen:2010yh} & (6360, 148) & (6440, 127)  & $\cdots$ & $\cdots$ \\
NST2~\cite{Nguyen:2010yh} & (5830, 320) & $\cdots$ & $\cdots$ & $\cdots$ \\
FKW~\cite{Fischer:2014cfa} & (6354, $\cdots$) & (6498, $\cdots$) & 6714 & $\cdots$ \\
G-RHK~\cite{Gomez-Rocha:2016cji} & (6275, $\cdots$) & (6334, $\cdots$) & & \\
HGKL2~\cite{Hilger:2017jti} & (6608, 306) & (6690, 273) & $\cdots$ & $\cdots$ \\
\hline \hline
   & \multicolumn{3}{c}{amplitudes}  \\
\hline\hline
$E_H$ & 1.408 & 0.276 & 0.027 & 0.015 \\
$F_H$ & 0.003 & $\cdots$  & $\cdots$ & $\cdots$\\
\hline\hline
\end{tabular}
\caption{\label{tab:mbc_all_opt} Mass spectrum of ground-state $B_c$ mesons.
  Our results were obtained with the best-fit parameter set:     $\alpha_{\mathrm{IR}}= 0.93\pi/59.046$ and $\Lambda_{\text{UV}}= 4.244\,\GeV$.
  $^*$The decay constants are from a recent Lattice calculation~\cite{Colquhoun:2015oha}.}
\end{center}
\end{table}
%
%
Finally, for the $B_c$ spectrum, we found $\hat\alpha_{IR}= 0.049$ and $\Lambda_{\text{UV}}= 4.244$, which are between the charm and bottom region according to the second and third column of \tab{tab:rc}.

We present our results in \tab{tab:mbc_all_opt} and see an excellent agreement with other model results in the case of the predicton for $B_c ^*$. Additionally, from all the predictions enlisted in that table, our CI model result is the closest of those predicted by quark models~\cite{Eichten:1980mw,Stanley:1980zm,Buchmuller:1980su,Godfrey:1985xj,Gershtein:1987jj,Kaidalov:1987gk,Kwong:1990am,Baker:1991ty,Chen:1992fq,Itoh:1992sd,Eichten:1994gt,Bagan:1994dy,Zeng:1994vj,Roncaglia:1994ex,Kiselev:1994rc,Gupta:1995ps,Fulcher:1998ka,Ebert:2002pp,Ikhdair:2003tt,Ikhdair:2003ry,Godfrey:2004ya,Ikhdair:2004hg,Ebert:2011jc}, 
light-front quark model (LFQM) \cite{Frederico:2002vs,Choi:2009ai,Choi:2015ywa}, 
reductions of the BSE (BSR) \cite{AbdElHady:1998kc,Baldicchi:2000cf,Ikhdair:2004tj},
with the nonrelativistic renormalization group (NRG) \cite{Penin:2004xi}, 
QCD sum rules (QCDSR) \cite{Aliev:1992vp,Gershtein:1994jw,Wang:2012kw}, and lattice QCD (LAT) \cite{Davies:1996gi,Chiu:2007bc,Gregory:2009hq,Dowdall:2012ab,Burch:2015pka}.

For this reason, we expect that our $B^*_c$ meson prediction to be in perfect agreement with a future experimental value. While for the $B_{c0}$ and $B_{c0}$ mesons we expect to be within to 2\% below future experimental data.

\section{\label{sec:conclusions} Conclusions}

We have extended the CI model, previously employed to calculate properties of light and heavy-quarkonia mesons, to study the mass spectrum and decay constants of heavy-light mesons. In the interest of study these systems, we used a fit inspired in a previous work ~\cite{Raya:2017ggu}. With this fit, we proposed and systematical scheme to calculate mesons in diverse regimes: we kept the dressed quark masses used in previous works, and then fixed the $\Lambda_{\mathrm {UV}}$ parameter to obtain the mass of $D$, $B$ and $B_c$ mesons and their respective coupling strength.

For charmed mesons, we found an $\alpha_{\mathrm{IR}}$ divided by a factor of 4.528 and increase $\Lambda_{\text{UV}}$ from 0.905\,\GeV to 1.532\,\GeV with respect to light quarks parameters, so our parameters are consistent with our inference about finding them in a region between the light and charm sector. Similar values are given in Ref.~\cite{Serna:2017nlr}. For $B$ and $B_s$ mesons, we divided $\alpha_{\mathrm{IR}}$ by a factor of 11.969 and increase $\Lambda_{\text{UV}}$ from 0.905\,\GeV to 2.223\,\GeV. Finally, to study $B_c$ mesons, we divided the coupling by a factor of 59.046 and increase $\Lambda_{\text{UV}}$ to 4.244, values between the charm and bottom parameters as we expected.

The mass spectrum for heavy light mesons is in excellent agreement with experimental data and other SDBSE results. For  charmed mesons we reproduce the experimental data within a 2\%. We can calculate values of scalar and axial-vector mesons, while due other SDBSE models struggles due to the really heavy computational calculations. For $B$ and $B_s$ mesons, our spectrum is again in agreement with experimental data. We predict a value for $B_0 (5610)$\,\MeV, which we expect to be around 60\,\MeV\, below the experimental one. We also predict a $B_{s0} (5701)$\,\MeV, which we expect to be around 75\,\MeV \, below experiment.

Additionally, we predicted a fully spectrum for bottom-charm mesons. If we consider the accuracy of our lighter mesons results, we claim that the mass of $B_{c}^* (6308)$ will be very close to experiment. While the masses of $B_{c0} (6490)$ and $B_{c0} (6518)$ will be below 2\% the future experimental measures. 

However, on attempting to calculate the decay constants for these unequal-mass systems, we found values that do not agree with experimental data, but neither other SDBSE results agree with them. This assures that studies beyond rainbow-ladder approximation are needed to a fully description of heavy-light mesons.  

Finally, with the results of this work and the recent CI model meson studies, we are able to mimic the high momentum tail of the quark mass function determined in the SDBSE studies on QCD~\cite{Maris:1999nt,Bhagwat:2004hn,Maris:1997hd}. Also, we find out that the reduction in the coupling model is dependent on the dressed quark masses involved in the bound-state. In order to compare this reduction with running coupling QCD, we fitted the contact interaction coupling with an inverse logarithmic curve. We expect that future works continue exploiting this feature to study baryons and exotics with a variety of quarks components.

This study is part of the series of studies on heavy-quarkonia in a contact interaction~\cite{Bedolla:2015mpa,Bedolla:2016yxq,Serna:2017nlr,Raya:2017ggu}, in which we move towards a comprehensive study of mesons and QCD by using this model. Further steps will involve baryons and exotics.

\section{Acknowledgments}

The authors acknowledge financial support from CONACyT, M\'exico (postdoctoral
fellowship for M.A.~Bedolla), the INFN Sezione di Genova and the Instituto de Fisica y Matematicas of Universidad Michoacana de San Nicolas de Hidalgo. We also acknowledge C. D. Roberts for his fruitful comments during the NPQCD18 in Seville, Spain.

\end{document}